\documentclass[nofootinbib,apsspec,showpacs,groupedaddressx,preprintnumbers]{revtex4-1}
\usepackage{amsmath}
\usepackage{amssymb}
\usepackage{graphics}
\usepackage[dvips]{graphicx}
\usepackage{graphicx}
\usepackage[usenames]{color}
\usepackage{wrapfig}

\newcommand \beq{\begin{eqnarray}}
\newcommand \eeq{\end{eqnarray}}
\def\simge{\mathrel{%
       \rlap{\raise 0.511ex \hbox{$>$}}{\lower 0.511ex \hbox{$\sim$}}}}
\def\simle{\mathrel{
       \rlap{\raise 0.511ex \hbox{$<$}}{\lower 0.511ex \hbox{$\sim$}}}}

\newcommand{\Slash}[1]{{\ooalign{\hfil$#1$\hfil\crcr\raise.167ex\hbox{/}}}}

\usepackage[colorlinks=true,linktocpage=true,linkcolor=blue,citecolor=blue]{hyperref}

\begin{document}
\title{Biographical Memoir of Stirling Colgate}

\author{W. David Arnett  }
\affiliation{Steward Observatory, University of Arizona, Tucson, Arizona, 85721}

\author{Gordon Baym }
\affiliation{Department of Physics, University of Illinois at Urbana-Champaign, 1110 W. Green Street, Urbana, Illinois 61801 \\  \\and}

\author{Necia (Nikki) Cooper}
\affiliation{Los Alamos National Laboratory (ret'd), Los Alamos, New Mexico 87545 \\ }
\date{\today}
\begin{abstract}

\end{abstract}

\begin{figure}[b]
\begin{center}
\includegraphics[scale=0.15]{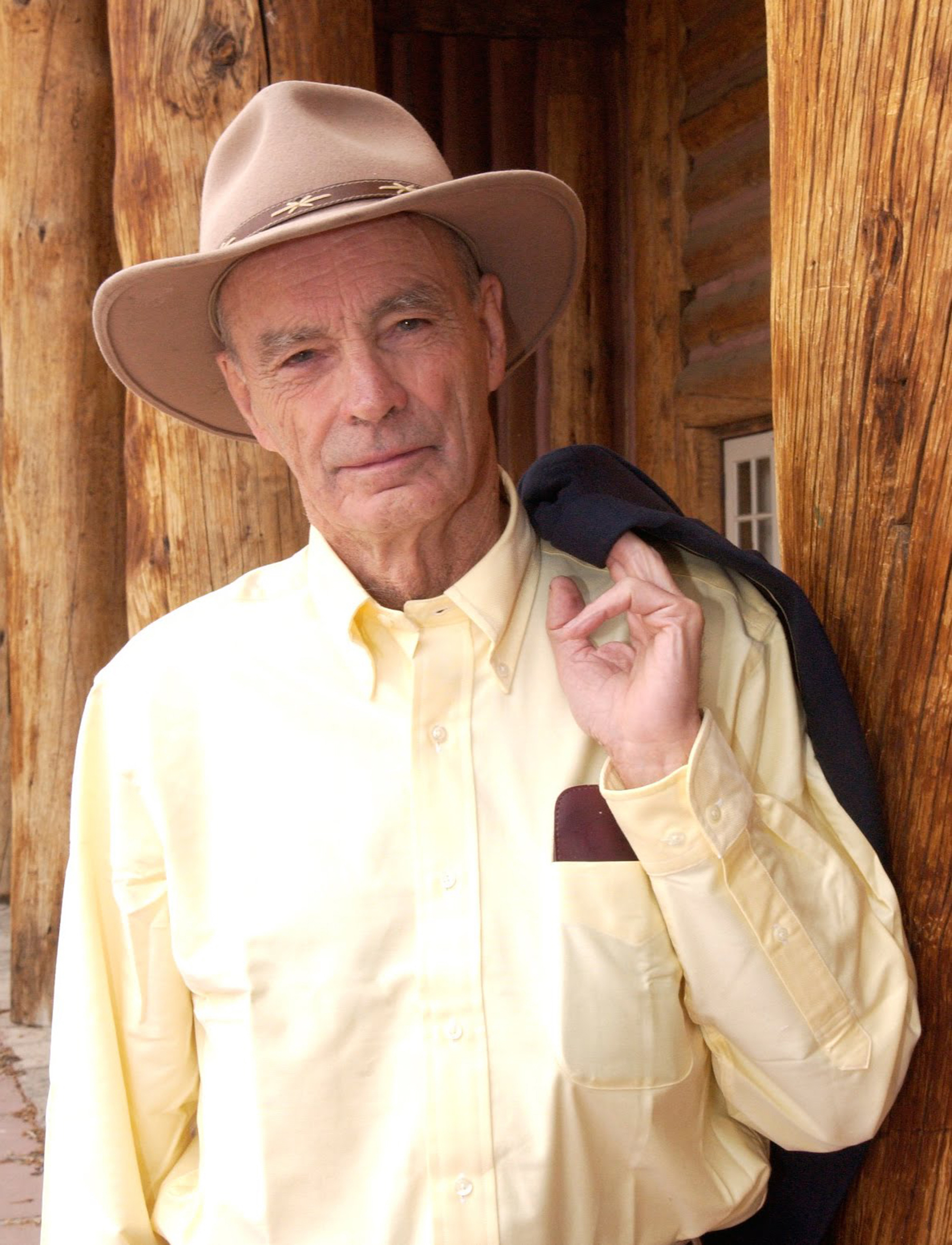}
\end{center}
\caption{\footnotesize{Stirling Colgate  (Los Alamos National Laboratory, courtesy AIP Emilio Segr\`e Visual Archives, Physics Today Collection).}}
\label{stirling-older}
\end{figure}

\maketitle

 Stirling Colgate was a remarkably imaginative physicist, an independent thinker with a wide breadth of interests and contagious enthusiasm, a born leader with enduring drive to attack fundamental problems in science.   Among his many achievements, he founded the quantitative theory of stellar collapse and supernova explosions, and introduced numerical simulation into the astrophysical toolbox.  He brought strong physical intuition to both theory and experiment, in the sciences of nuclear weapons, magnetic and inertial fusion, as well as astrophysics.

\section{Early years}

     
  Born Stirling Auchincloss Colgate on Nov. 14, 1925, in New York City, to Henry Auchincloss Colgate (of toothpaste company fame) and Jeanette Thurber (n{\'e}e Pruyn) Colgate, he attended and graduated from the Los Alamos Ranch School.   This is the place General Leslie Groves and Robert Oppenheimer chose in the fall of 1942 to be the initial site of the Los Alamos Laboratory of the Manhattan Project.  When Oppenheimer made an incognito visit to the Ranch School with Ernest O. Lawrence, Stirling in fact recognized them as famous nuclear scientists and suspected that the school was being taken over to build an atomic bomb!\footnote{As Stirling recalled, ``I was a seventeen-year-old student at the Los Alamos Ranch School in 1942.
I remember when the bulldozers came through to remake the school. About
December that year, two men showed up at school, and we were required to say
our yes sirs to a Mr. Jones, who was wearing a fedora, and to a Mr. Smith, who
was wearing a porkpie hat. The names were obviously pseudonyms. Not only
was everybody showing them great deference, but Mr. Jones seemed most uncomfortable every time someone referred to him by that name.

The four of us who were seniors had studied physics. The pictures in our
physics textbook made it easy for us to recognize Mr. Jones as Ernest Lawrence
and Mr. Smith as Robert Oppenheimer. Furthermore, the discovery of fission
had been big news. In fact, we were even aware of the idea of a chain reaction.
Clearly, the school was about to be converted to a laboratory to work on a
very secret physics project. Why else would top physicists be visiting a place
out at the end of nowhere with no water, no roads, no facilities? What was really going on was obvious! We were secretly amused by the pretense"  \cite{teller}.}

 \begin{wrapfigure}[24]{R}{5cm}
 \vspace{-12pt}
\begin{center}
\includegraphics[width=5cm]{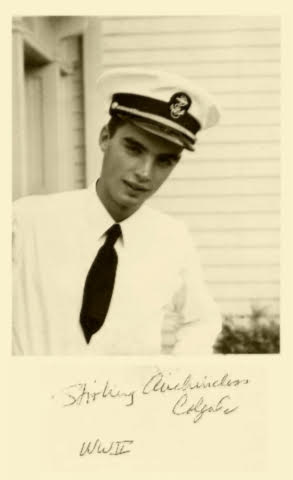}
\end{center}
\vspace{6pt}
\caption{\footnotesize{In the Merchant Marine, WWII}}
\label{merchantmarine}
\end{wrapfigure}

  Keeping this secret, Stirling returned East in 1943 to enter the electrical engineering program at Cornell University, but a year later joined the U.S. Merchant Marine to participate in the war effort (1944-46).  He was in the Pacific on a merchant ship staffed by a Dutch crew when the Hiroshima bomb was dropped, and at the captain's request shared with the ship's officers and crew his ideas of how a nuclear fission bomb works.  The captain was so impressed that he ``ordered'' Stirling to study physics after completing his service.

   Indeed, on his return to Cornell in Sept 1946, he changed his major and earned an A.B. in physics in 1948 and then a Ph.D. in experimental nuclear physics in 1952 under Robert R. Wilson.  His thesis project, which would serve him well in his future career, was to fulfill a request from the National Bureau of Standards for high-accuracy measurements of absorption and scattering rates of  gamma rays of varying energy in various materials, which he did using NaI scintillation counters \cite{thesis}.  

\section{Early Career}

 In January 1952 Stirling began a postdoc with Luis Alvarez at Berkeley, where he conceived and successfully tested an accelerator 
designed to inject a very high current, low energy deuteron beam into an Alvarez-designed linear accelerator for a short-lived effort to use accelerator-driven neutrons for producing plutonium from U-238.   
However, in
late summer1952 Stirling left his postdoc position, lured to the newly created Livermore National Laboratory by Director Herb York expressly to participate in Project Sherwood, an Atomic Energy Commission-sponsored, secret program to control magnetic-confinement fusion. 

  Almost immediately, York and Edward Teller asked Stirling to postpone work on fusion and instead help the Los Alamos Laboratory's nuclear weapons program by designing the diagnostics for the reaction history of the upcoming 1954 thermonuclear weapon (Castle Bravo) test at Bikini atoll, 
 as well as to design similar diagnostics for two planned Livermore shots.    Stirling and his group dramatically improved the techniques for measuring the intensity of the prompt neutrons and gamma rays over time from multiple locations on the device. 
 
\begin{wrapfigure}[20]{L}{6cm}
\begin{center}
\includegraphics[width=6cm]{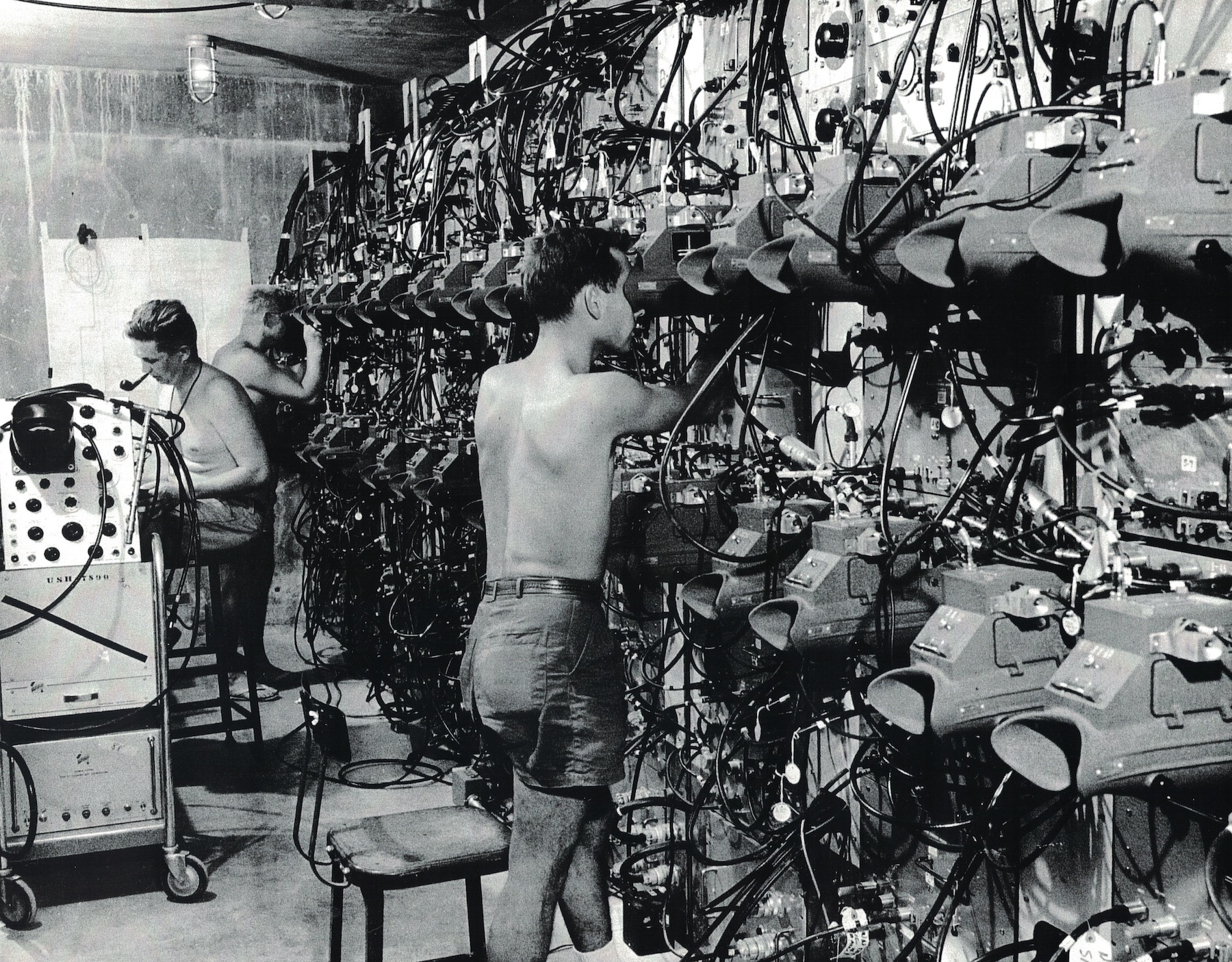}
\end{center}
\vspace{6pt}
\caption{\footnotesize{Stirling checking the oscillograph cameras in the recording bunker at Bikini, Feb. 11, 1954.}}
\label{bravo}
\end{wrapfigure}

   At the outset Stirling consulted with senior scientists Marshall Rosenbluth and Conrad Longmire at Los Alamos to learn how and why the Bravo device was designed the way it was and what diagnostics would reveal the reasons for failure should the yield be much less than predicted.   This insistence on understanding the physics, combined with Stirling's exceptional ability to design experiments proved to be essential to the success of the diagnostics.   To start, he and his team found at Bikini that light from a test lamp would not get all the way through the two kilometer pipeline to carry the signal from the device to the recording bunker (Fig.~\ref{bravo}).   He realized immediately that the contractors, forgetting that the earth is round, made the pipelines locally level with the earth rather than straight, a problem soon corrected.

   Stirling's second and bigger save came a week prior to the shot, when he was shown photos of a test from the Nevada Test Site, with luminous jets flowing along guy wires ahead of the fireball.    When no one from Teller on down could offer Stirling a convincing explanation for this puzzling phenomenon, Stirling estimated that the energy a similar but scaled-up jet flowing along the steel pipelines would deliver to the recording bunker would be the equivalent of one to two kilotons of TNT, likely enough to obliterate the bunker and its contents. 
He then recommended that 100,000 tons of coral sand be piled up ahead and on top of the recording bunker to blunt the impact.     
Indeed, after the Bravo shot, most of the coral was gone, and one of the bunker doors was ajar, but the data were saved; key parameters including the rate of thermonuclear burn, measured to high accuracy, would 
become a benchmark for computer simulations of device performance for years to come \cite{Ogle}.

   In the summer of 1954, after completing the shot reports for Bravo, Stirling was free to join Project Sherwood, an opportunity for free-thinking that would lead to his lifelong involvement with supernova explosions and other high energy astrophysical phenomena. 
He went, as he would say, ``full bore'' into learning the theory and designing two new experiments related to magnetic pinches, the simplest method, in principle, to achieve controlled thermonuclear fusion. 

 \begin{wrapfigure}[20]{R}{6cm}
\begin{center}
\includegraphics[width=6cm]{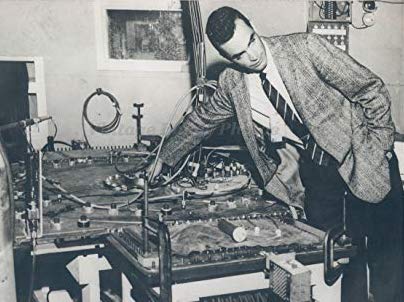}
\end{center}
\vspace{6pt}
\caption{\footnotesize{Explaining the dynamical pinch at Livermore, 1958.}}
\label{explaining}
\end{wrapfigure}

In 1955 the Berkeley linear 
Z-pinch was producing large neutron yields on a regular basis, and hopes were high that the neutrons were originating from thermonuclear fusion.  With characteristic physical insight and analysis, Stirling debunked this conclusion and saved the program from embarrassment, showing that the plasma temperature was much too low to produce a large thermonuclear neutron yield within the observed short duration of the neutron pulse.
He then led the Berkeley team in designing a nuclear emulsion experiment that would reveal unequivocally that the pinch neutrons arose from fusion reactions initiated by``accelerated'' deuterons, not from thermonuclear fusion. 
and showed that the acceleration was driven by
an instability earlier predicted in 1954.
All the elements fitted together quantitatively, and his explanation was soon accepted
by the entire pinch community \cite{pinch}.\footnote{His further work on stabilizing pinches culminated in Rosenbluth's rigorously developing in 1956 a necessary (but not sufficient) condition for pinch stability, which lead the U.S. and U.K. pinch communities to modify their machines.    An immediate outcome, observed by Stirling and Harold F\"urth, was a ``transient interval of stability, lasting some ten times longer than the life-time of the pinch when unstabilized'' \cite{furth}.    Further experiments by Stirling and coworkers then revealed the all important filamentation, or resistive tearing, mode, which plays a major role in both astrophysical plasmas and the operation of the Tokamak and the Reversed Field Pinch devices \cite{furth2}.  }  Overall Stirling's early research was a seminal influence on the design of
devices for producing controlled thermonuclear fusion. Furthermore, his firsthand
experience with the behavior of laboratory plasmas led to his understanding
the generation of cosmic rays in solar flares and radio lobes \cite{cosmic}.

    In the meantime, Teller, concerned in 1956 with detecting potential US and Soviet high altitude nuclear explosions, told Stirling to think about the signals from thermonuclear explosions in vacuum.   Back in 1953-54, while calculating all the background radiation signals relevant to the Bravo reaction history measurements, Stirling started exploring whether shock breakout through the steep density gradient created by Bravo might create a new X-ray and gamma ray background signal never before considered.   In vacuum that effect would be much larger, so in 1956 Stirling, with Montgomery H. Johnson, began predicting the magnitude of this effect for a hydrogen bomb explosion in space.  At the same time, they applied this idea to shock breakout in supernova explosions, -- in which the core of a star at the end of its evolution collapses generating an outward moving shock -- to explore whether that process might be a significant source of radiation from space, including the observed cosmic rays. This track would lead to a thoroughly unexpected major scientific discovery.    
    
Recruited by the U.S. State Department in 1959 as the scientific consultant on nuclear test ban negotiations in Geneva, Stirling learned that the negotiations were at an impasse -- the Soviets wanted a nuclear test ban based on trust, rather than intrusive surveillance. 
The head of the U.S. delegation told Stirling that the U.S. government, as a way to reach agreement with the Soviets, wanted to use satellites to detect nuclear testing and verify compliance with a test ban.  
Stirling immediately realized the relevance of the shock-breakout radiation signals from both thermonuclear explosions and supernovae,  for the test ban negotiations.  As he described events in 1959, 
``By then I had \dots estimated that a gamma-ray signal from a supernova would look similar to the gamma-ray signal from a nuclear bomb blast in space.  So I gave a talk about my research, pointed out the weakness of their proposed monitoring system, and asked the question: What if someone mistook a supernova for a nuclear bomb? There was great consternation among the Soviet delegation, I mean real consternation. This idea caught them totally by surprise. After 20 minutes of huddling, the Soviets responded, `Who knows what a supernova will do? We'll have a two-week recess to consider all of this.'  Of course my immediate reaction was, `I'm going to show you what a supernova is!' and that, of course, is how I made my career in astrophysics" \cite{SC-LLNL}.
 
After the recess, to the surprise of Stirling and the U.S. team, the Soviets agreed to help develop a satellite system that could differentiate signals coming from inside the solar system versus those from outer space. The U.S. immediately started developing the series of Vela satellites. Four years later the Air Force launched the first pair of Velas on October 17, 1963, two months after Kennedy and Khrushchev signed the Partial Test Ban Treaty and one week after the treaty went into effect. 

   Los Alamos scientists were tasked with building the radiation detectors and analyzing the data, which were classified.   Stirling and Teller urged them to look for correlations between bursts of gamma rays and new optically-observed supernovae, but initially no such correlation was found. However, the better instrumentation on the upgraded Vela 5A, 5B, 6A, and 6B satellites allowed them to peer deeper into space and record rarer events, and thus in the years 1969--1972 Los Alamos scientists discovered 16 cosmic gamma ray bursts coming from outside the solar system, each lasting several seconds and dominating the gamma radiation of the entire sky. In 1967, Stirling wrote up his model for gamma-ray bursts presented at the 1959 test ban negotiations, concluding that shock breakout through the surface of a supernova explosion from a collapsing star would cause a burst of X-rays and gamma rays detectable by the Vela satellites \cite{SC1968}.   He updated the model in 1974 \cite{earlygamma} after Los Alamos published their discovery of gamma-ray bursts in 1973 \cite{klebesadel}.   As we know now, an important subset of such bursts, the long bursts, are indeed associated with distant core collapse supernovae.
  
   Stirling's early research on supernova shock breakout as a source of cosmic rays was published with Johnson \cite{cj-cosmic1960}. In this paper, Stirling lays out plans to model the entire supernova explosion from nuclear disassociation and the onset of instability to the blowoff of the outer layers, determined, no doubt, to show the Soviets ``what a supernova will do.''

 \begin{wrapfigure}[16]{t}{6cm}
 \vspace{-24pt}
\begin{center}
\includegraphics[width=6cm]{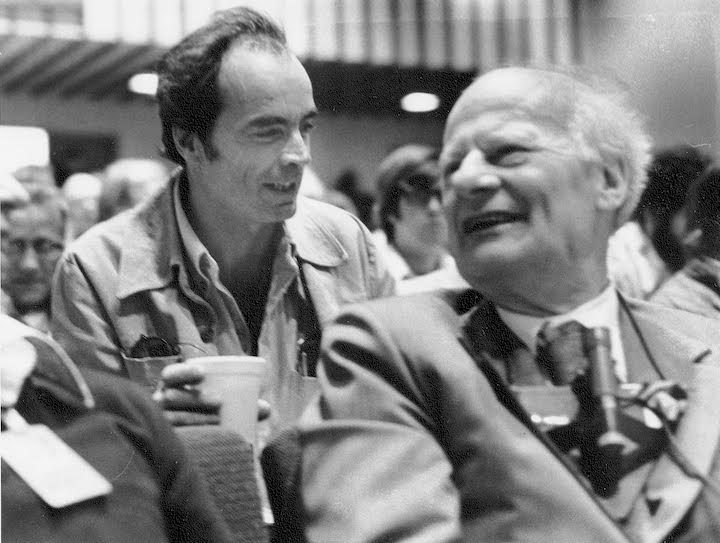}
\end{center}
\caption{\footnotesize{With Hans Bethe.}}
\label{stirling-hans}
\end{wrapfigure}

   Then, in 1961, Stirling, William H. Grasberger, and Richard H. White, did the first numerical simulation of stellar collapse in a supernova \cite{grasberger}.   This work started the field of quantitative modeling of supernova explosions, and was important as one of the first introductions of time-dependent numerical hydrodynamic calculations into astrophysics.  Their first modeling led to total collapse without explosion, which we would interpret now as the formation of a black hole.  Realizing the importance of making a neutron star in the supernova, they tried stiffer equations of state and succeeded in generating a ``bounce," the rebound of the collapsing core, with formation of a shock wave, albeit with an inadequate explosion.  They did not, however, take into account at the time neutrino processes in supernovae, a key ingredient that Stirling would later introduce.

  Still searching for the correct physics of supernova explosions, and taking advantage of the computing power and software available at Livermore, Stirling and White numerically simulated the 1964 Fowler-Hoyle model of a core collapse supernovae  in which the energy driving the explosion was taken to be nuclear burning of $^{16}$O \citep{fowler-hoyle}.   Their simulations showed that burning was unable to make the star explode; rather, the entire star collapsed \cite{CW1964}.  
  
   Fowler and Hoyle had assumed that contraction of the star might be held in check by its rotation, an option that could not be explored by the one-dimensional  codes of the time.  Then in 1965 Colgate and White presented the ingenious and prescient idea that neutrinos emitted in the core collapse could transport sufficient energy to explode the mantle \cite{CW1966}.\footnote{First written up in April 1965 as a Livermore preprint \cite{cwpreprint}, the publication of the Colgate-White paper was delayed in part by the referee; S. Chandrasekhar, then editor of the Astrophysical Journal, eventually overruled the referee and accepted the paper. As a consequence, Cameron advised Arnett to publish his dissertation and related work in the Canadian Journal of Physics because of such refereeing issues.}$^,$\footnote{In 1988 when Hans Bethe was writing his review of supernova mechanisms \cite{HAB},  Stirling wrote him to offer his own perspective on his supernova work \cite{SACtoHAB}, especially the 1966 paper: ``Several aspects of this paper \dots are
particularly important to what happened later. The verification of the
numerical hydrodynamic code with the test problems of a shock in a
density gradient, the free-fall gravity solution and the test that a
star modeled with this code exhibited both the phenomena of stability
and subsequent collapse sensitive to the equation of state. The
inadequacy of the bounce shock in a polytropic structure of index three
was evident so that the formation of a neutron star became the issue.   
Without the neutral current cross section for neutrinos unknown [sic] then,
the collapse to a neutron star produced a deleptonized neutron star with
all the neutrinos escaping at relatively low energy, and hence, small
cross section. The deleptonization using the thermally free protons,
although only briefly mentioned, took significant understanding. Later
the approximation of an accretion shock giving rise to the neutrino flux
creating mass ejection is of course the current still unresolved issue.  
What was more puzzling to me at that time was given a neutrino derived
explosion, how could one create a reasonable optical supernova [light curve].''}
Carrying over their experience with weapons, they envisaged that neutrino radiation would be similar to photon radiation in a bomb, and built a simple model  in which they assumed that
 the neutrinos would dump half of the gravitational binding energy released in the core into the mantle, invariably producing an explosion and a neutron star. 
 They concluded (owing to errors in the thermal component of the assumed equation of state) that the shock structure would lead only to neutron stars, and not black holes \cite{CW1966,colgate68}.   An improved equation of state gave both black holes and neutron stars, depending on core mass \cite{wda66} and neutrino thermalization \cite{wda68}.\footnote{As a result of the space race,  the new Goddard Institute for Space Studies had in 1964 an IBM 360/95 computer which was more powerful and less clogged than the best computers then available at LLNL and LANL, allowing Arnett to develop from scratch an ``outside the fence" hydrodynamics code with radiative diffusion (the lab codes were then top secret; Stirling's notes to Cameron, Arnett's dissertation advisor, and Christy's unclassified article \cite{christy} were helpful clues, as were the test problems provided in \cite{CW1966}.}.

    Stirling responded to Arnett's discovery of possible black-hole formation in core collapse with a denial, writing on the difference of their and Arnett's treatment \cite{wda66,wda67} of neutrino transport,  ``Conversely, all stars undergoing non-relativistic collapse according to the mechanisms of CW may manage to eject sufficient mass so the residual neutron star is stable" \cite{colgate68}.  But later, in his 1988 letter to Bethe  \cite{SACtoHAB}, Stirling would acknowledge Arnett, writing, ``The inadequacy of the thermonuclear energy for creating an explosion was also calculated. I did not at that time recognize what Dave Arnett later saw that for a small mass star a thermonuclear explosion is adequate to cause mass ejection as in Type I's.''
 
 Detection of approximately thermalized neutrinos from the supernova SN1987A  showed that Stirling's intuition on the importance of neutrinos was indeed correct.

\section{From Livermore to New Mexico Tech and then Los Alamos}   

   Stirling left Livermore in 1965 to become Professor of Physics and then President at the New Mexico Institute of Mining and Technology (New Mexico Tech) in Socorro, from 1965-1974.         He came to Los Alamos National Laboratory (LANL) as a full time staff member in 1976, becoming a Senior Fellow in 1982 and Senior Lab Fellow in 1987; he was a Lab Associate Fellow until his death in 2013.

   In 1968, Stirling went on to analyze the light emitted by supernovae over time (the ``supernova light curve"), a problem harking back to his earlier work on detecting products of thermonuclear explosions.    Taking into account the new prediction of nucleosynthesis of radioactive $^{56}$Ni during silicon burning, the last thermonuclear gasp of a dying star \cite{truran}, he and Chester McKee produced the first predictions of supernova light curves powered by $^{56}$Ni decay \cite{mckee}. They point out that their predictions fit Type-I supernova particularly well, and emphasize that their analysis is not tied to the particular explosion mechanism, but rather that for all supernovae, ``roughly 1 solar mass is ejected with a mean velocity corresponding to the gravitational binding energy just before explosion and with a velocity distribution depending on the relative location of the mass fraction in question.''    In 1979 when interest arose in using distant supernova as a standard candle to measure the Hubble constant and the deceleration parameter $q_0$ of the universe, Stirling outlined how a coordinated program using Type I supernovae, with their high intensity and relatively consistent light curves, as the standard candle at high redshift would determine the acceleration parameter with greater accuracy than other possible standard candles \cite{candle}.

     Over the years Stirling remained focused on the supernova problem, particularly on finding the mechanism that would reliably produce an explosion with the observed kinetic energy of a few times 10$^{51}$ ergs.  Realizing from weapons work that multi-dimensional asymmetries upon compression would grow, and that the same problem would carry over to supernova collapse,  he emphasized the need for multi-dimensional simulations to correctly account for non-spherical instabilities in core collapse, and in particular neutrino convection \cite{convection,epstein}.  In 1992 he gave crucial guidance to Willy Benz and his then graduate student Marc Herant, both from the Center for Astrophysics and Harvard, redirecting their focus from late-time (300 seconds after collapse) multi-dimensional hydrodynamics to simulating the first few 
\newpage
 \noindent seconds  after core-collapse as a way to explain the large abundance and high velocities of $^{56}$Ni and its decay products in the SN1987A light curve, and the large asymmetries observed in the SN1987A expanding envelope. Stirling emphasized the need to start from a high entropy bubble interior to the relatively low entropy matter behind the shock, as in the ``delayed explosion" mechanism of Wilson and Mayle \cite{WilsonMayle}. Then, with Herant taking the lead they successfully developed a more realistic convection-driven explosion mechanism for core collapse supernovae \cite{herant}.  Bethe agreed with Stirling that this model solved the longstanding problem of determining a robust, self-regulating explosion mechanism, unresolved details not withstanding. 
  
       Stirling also realized from an early stage the importance of observing supernovae, and in 1971 proposed building a telescope for automated supernovae searches.  By 1975, when he was leaving New Mexico Tech to join Los Alamos, Stirling and colleagues had successfully designed and put into operation a remote controlled, fully automated 30 inch telescope for supernova searches, mounted at 10,000 feet in the Magdalena Mountains west of New Mexico Tech \cite{searches} -- the ``dig-as" telescope, as he called it.       Although the project  was plagued by pattern recognition
\begin{wrapfigure}[22]{t}{5cm}
\begin{center}
\includegraphics[width=5cm]{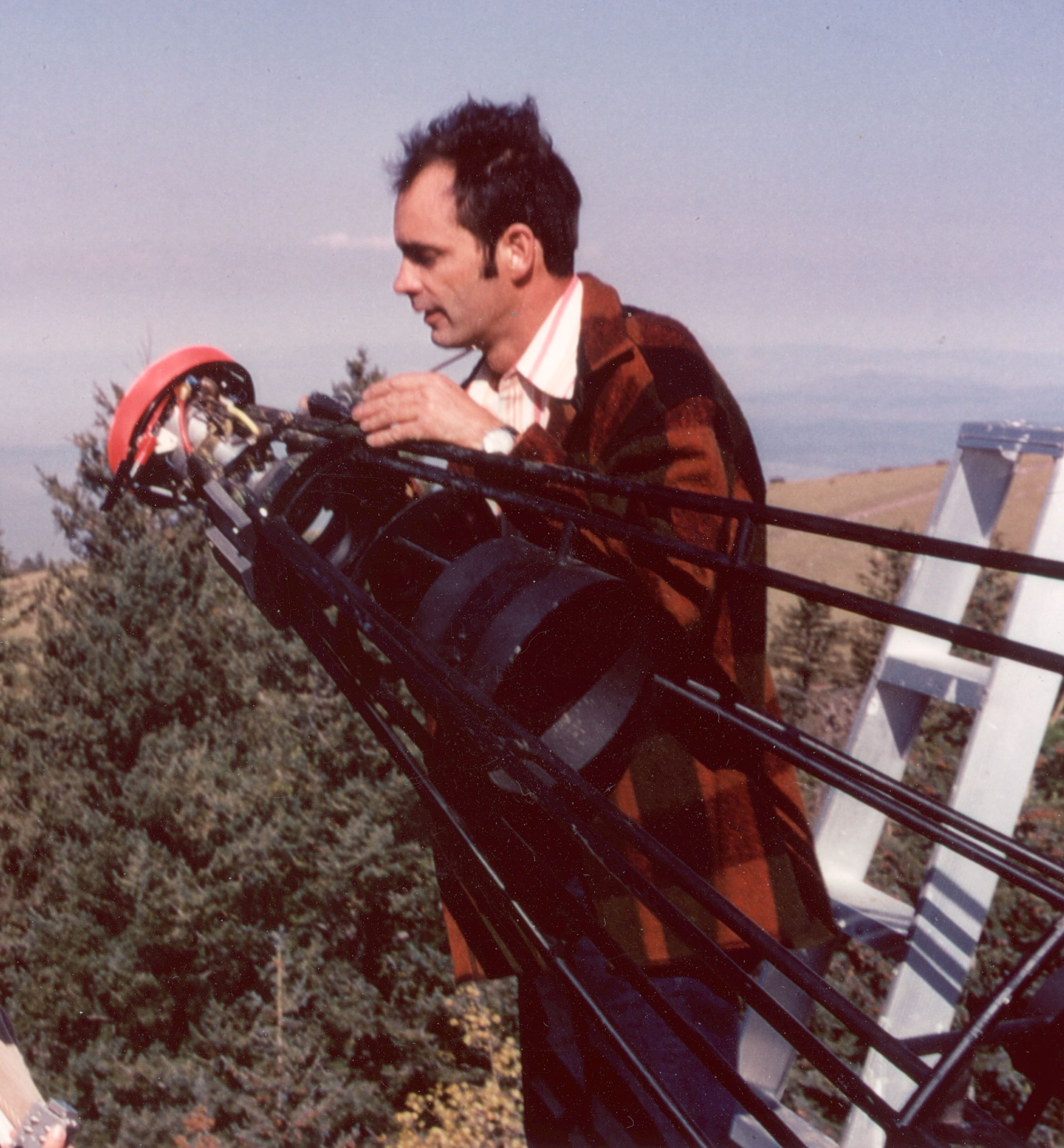}
\end{center}
\caption{\footnotesize{Adjusting the 30 inch telescope, ca. 1980.}}
\label{sSAC-mtn}
\end{wrapfigure}
 problems owing to inadequate technology, Stirling had planted the idea, and his impact on the building of a multitude of supernova search telescopes was clear.  In 1988, years after the 30-inch telescope was shut down,  Stirling spent time with Saul Perlmutter, who was starting a new automated supernova search, sharing his insights, experiences, and recommendations on how to implement an effective search.  Perlmutter and his team went on to discover the acceleration of the universe and dark energy; they write, ``Over the last decade, based on ideas of Stirling Colgate \dots we have developed the capability to search over 600 galaxies per night for supernovae'' \cite{pennypacker}.
And Perlmutter recounts in his 2011 Nobel Prize acceptance speech, "Luis Alvarez suggests to Rich Muller [Perlmutter's Ph.D. advisor] that it is time to re-do Stirling Colgate's robotic SN search" \cite{perlmutter},

    The origin of cosmic rays, especially of the newly discovered ultra-high-energy TeV cosmic rays, remained over the years among the  problems Stirling kept in mind; in an unpublished paper \cite{cosmicrays} he proposed that cosmic rays would originate in the huge radio jets emanating from active galaxies as they accrete material onto their central supermassive black holes.  That radio jets associated with active galaxies emit cosmic rays was recently confirmed in the TXS Blazar multi-messenger  discovery \cite{txs}.
His posited acceleration mechanism was through electric fields parallel to the magnetic fields that resulted from 
reconnection of force-free magnetic fields \cite{kronberg}.  
As was typical, he suggested that ``Laboratory experiments be performed to simulate both magneto-hydrodynamics as well as the tearing mode reconnection and the associated $E_\parallel$ acceleration of the `runaway' particles. \dots Interruptions in tokamaks are already laboratory proof of this acceleration''
\cite{cosmicrays}.

   Stirling was known for keeping lit the torch of experimental science in both astrophysics and nuclear weapons physics despite the increasing dominance of computer simulation in these fields. Until two years before his death he was flying back and forth from Los Alamos to Socorro in his own single-engine STOL-equipped Cessna 210 to work on the liquid-sodium astrophysical dynamo experiment  -- a project he designed to test his deep conviction and his and colleagues' theoretical work on the origin of cosmological magnetic fields.   This experiment at the limit of experimental technique, carried out with the help of graduate and undergraduate students at New Mexico Tech, was based on semi-coherent motions and run at the highest Reynolds and magnetic Reynolds numbers of all such dynamo experiments.   It succeeded in amplifying the seed field some eight times in the toroidal direction, the only experiment to produce significant field amplification \cite{magneticshear}.\footnote{Stirling's son Arthur Colgate, an industrial engineer,    
is currently implementing the second phase of the experiment, examining whether creating plumes in the rotating disc of sodium converts the toroidal field to radial, thus amplifying the seed field and completing the dynamo cycle.}

\section{Living and Sharing  His Passion for Physics}

\vspace{.25in}
      
      Stirling was a very colorful personality, always ready with a new idea and taking on projects that no one else would attempt.   A remarkable example was his suggestion, during the volcanic eruption on the island of Heimaey in Iceland in 1973, on how to prevent the lava flowing into the sea from blocking the narrow entrance to the harbor of Vestmannaeyjar, the most valuable seaport of the Islandic fishing industry. His plan was to detonate explosives at the lava-seawater interface near the harbor entrance, thereby inducing rapid water-lava mixing that would cool the lava and create a thick solid barrier that would impede and divert the flow.   The steps toward execution, ``with the help of the Icelandic government, the Icelandic Coast Guard and the U.S. Navy," were fully in progress when on further reflection Stirling realized the day before the detonation ``the awesome possibility that once mixing was initiated it might be self sustaining in that the high pressure steam produced might cause further mixing until all the lava had exchanged its heat with the water above it.   The energy released might have come to between 2 and 4 megatons. Naturally the experiment was calIed off."  \cite{heimaey-nature,nytimes}    Later, when he joined Los Alamos he initiated reactor safety research on possible auto-catalytic fluid-fluid-mixing explosions during reactor accidents, and raised the need for safety precautions to prevent similar auto-catalytic explosions arising from leaks in liquid natural gas tankers.

\begin{wrapfigure}[28]{L}{6cm}
\begin{center}
\includegraphics[width=6cm]{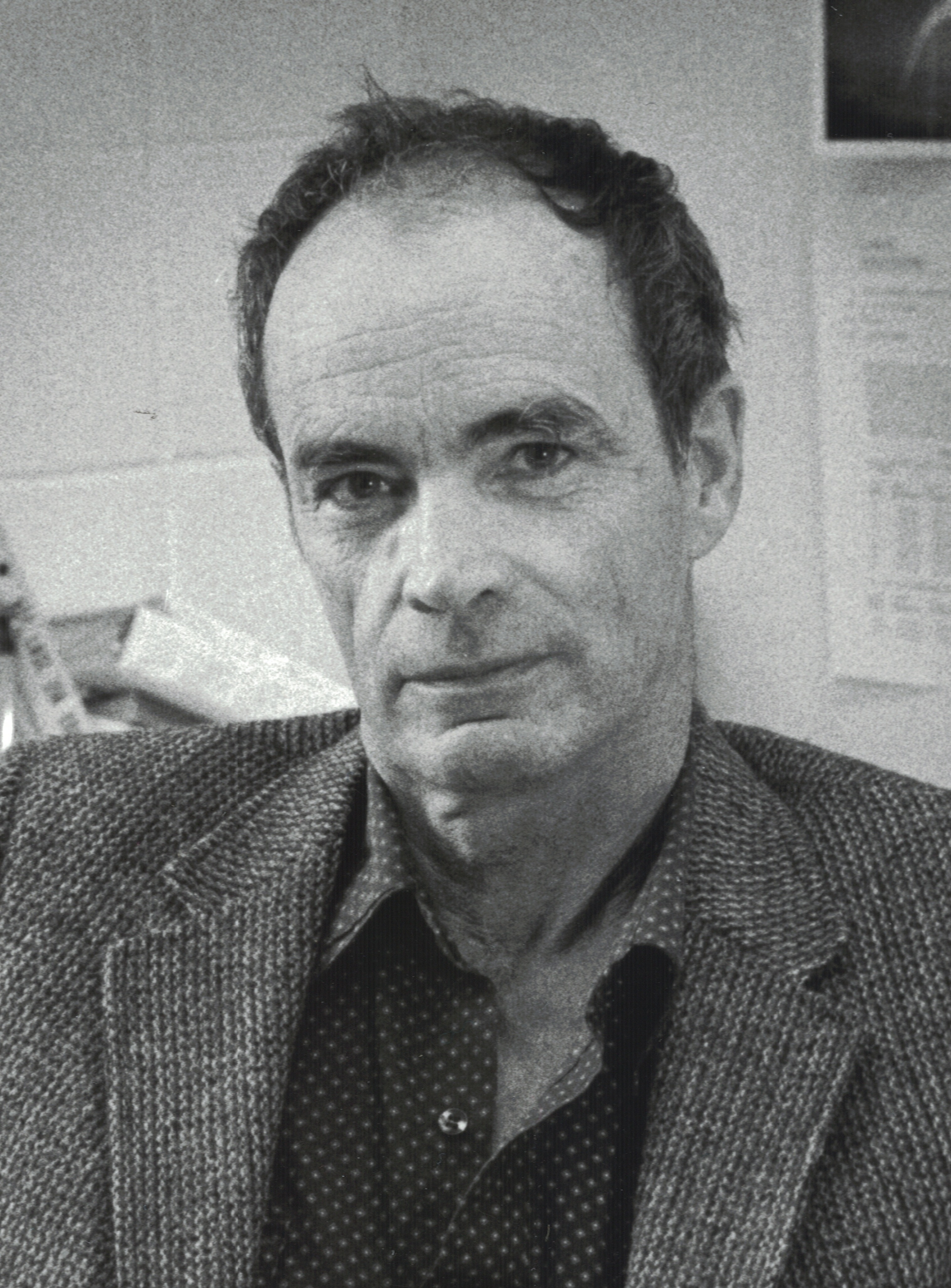}
\end{center}
\caption{\footnotesize{At Los Alamos National Laboratory, 1990.}}
\label{1990portait}
\end{wrapfigure}

    In 1981 when the first AIDS cases were publicized, Stirling was among those who recognized that the disease was about to turn into a world-wide epidemic.  He raised the consciousness of the Los Alamos Laboratory management as to the severity of the public-health threat and over time helped to get Laboratory support for a major multi-group research effort in the Lab's Theoretical Division to understand and help stem the epidemic.  Stirling himself pioneered the understanding that the distribution of risk-based behavior in different sub-populations was a major driver in the rate of transmission of the disease, and pointed out that the observed sub-exponential growth rate of the epidemic was not a sign that the programs to reduce its growth were succeeding, but rather that the slower growth was in the nature of the disease, and success was still far away.   With scientists from the Applied Math group he developed a risk-based model of the growth of AIDS in the U.S. based on the particular pattern of disease transmission.   Although Stirling did not stay in this field, he helped to shape the language of the modern debate on the spread of the disease \cite{aids}.

  As President of New Mexico Tech, Stirling found ingenious ways to integrate his administrative duties and research interests, remarking, e.g., ``It was easier to raise money for scientific research proposals than trying to raise it, as most presidents do, by gifts.''  He could then employ undergraduates in research jobs as other universities do with graduate students.  He also collaborated with the faculty and students in modernizing the curriculum, including for example a course on information theory that started with Godel's incompleteness theorem and went through Shannon's limit on the rate of information transfer. During his 10 years as president, the student body grew from 300 to 1100 of which 10 percent were graduate students, and 40 percent held 60 percent of the jobs on campus.   At the same time he continued his research activities in astrophysics, from supernova light curves to quasars to gamma-ray bursts, while expanding into the atmospheric physics of thunderstorms and tornados, and building a digitized telescope.  As President he published some three dozen research papers.
    \begin{wrapfigure}[18]{t}{5cm}
\begin{center}
\includegraphics[width=5cm]{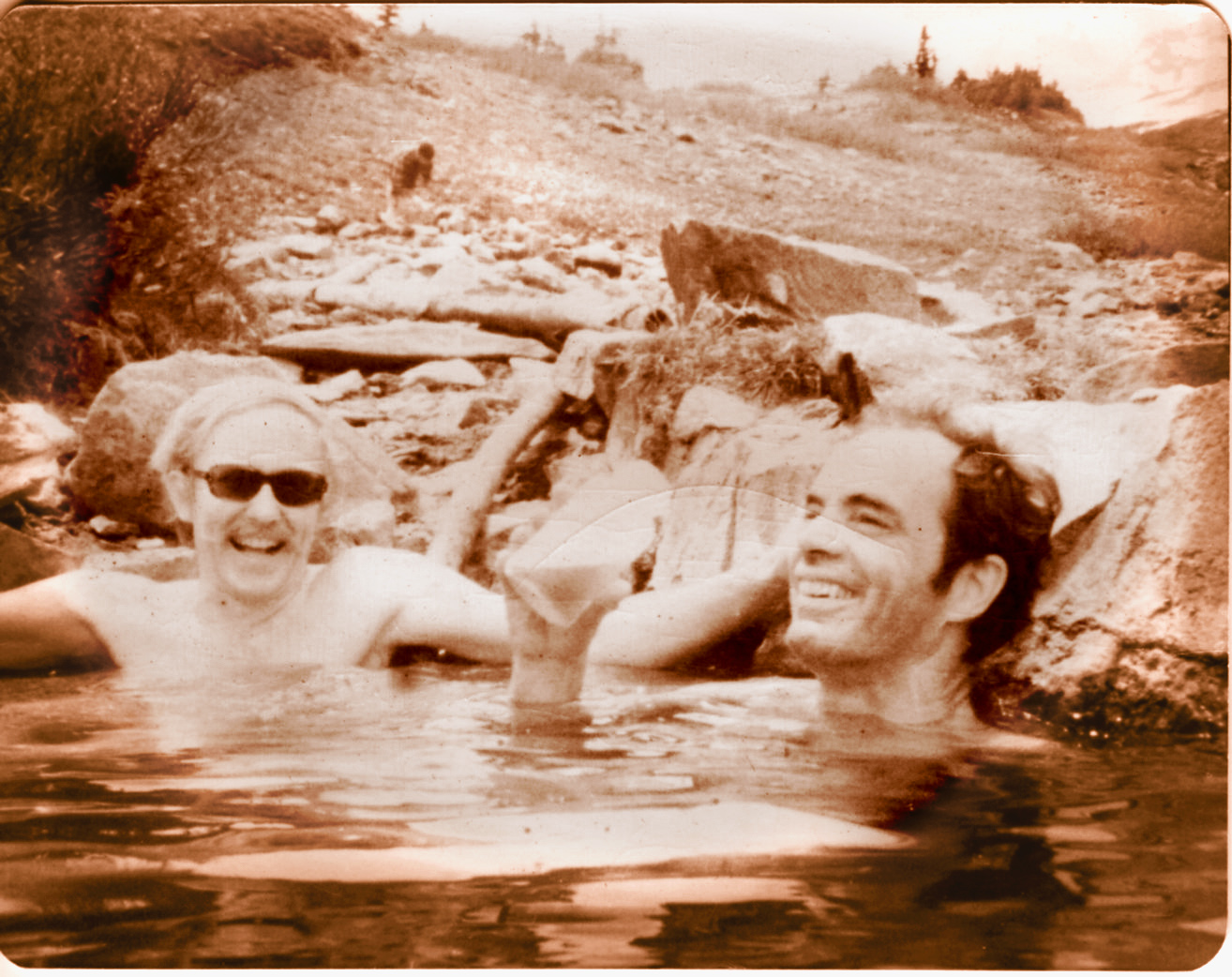}
\end{center}
\caption{\footnotesize{Stirling and Craig Wheeler at Conundrum Hot Springs, near Aspen, 1977.}}
\label{wheeler}
\end{wrapfigure} 

  When Stirling came to Los Alamos as a full time staff member in 1976, he was asked by then Lab Director Harold Agnew to form and lead a new group in Theoretical Astrophysics.  This new group attracted a large number of very able university scientists;   simultaneously Stirling and Al Cameron, developed a special postdoc program in which participants worked on both theoretical astrophysics and applied projects for the weapons program. Stirling himself served as a superb role model, contributing new ideas and creative approaches to both throughout his long career.  ``He has expanded his astrophysics research in myriad directions, always at the cutting edge and involving younger scientists whom he could mentor,'' commented Laboratory Fellow Johndale Solem in an internal letter recommending Stirling for the Los Alamos Medal.  Those new directions included galactic accretion disks, the galactic dynamo, the role of magnetic fields in the universe, the origin of cosmic radiation, and planetary formation.

  Stirling totally enjoyed communicating his love of science, a trait greatly admired by his students, and he was very supportive of younger people.    ``I think Stirling was a fantastic mentor to students and postdocs,'' said Los Alamos Theoretical Division Leader Tony Redondo. ``His office was always full of young people who were very excited to have discussions with him. Stirling always had very interesting ideas''   \cite{la-obit}.  Laboratory Fellow James Smith of the Materials Technology/Metallurgy Group mentioned, ``he really loved having young people around who wanted to talk to him. He really cared about helping young people''  \cite{la-obit}. Former student Dave Lee Summers wrote, "He taught me to always ask why things work and not just how they work" \cite{summers}.

   
   Stirling remained an intellectual leader at Los Alamos, holding people together and exhibiting rare compassion, which his colleagues thought stemmed from his true passion for physics. He was legendary for his insistence on understanding and his masterful ability to cut through the clutter of complicated details and find a simple explanation or critical link uniting the entire dynamics of a complicated system.  Theoretical plasma physicist Pat Diamond at the University of California San Diego and a close colleague of Rosenbluth, commented at his talk at the 2014 memorial symposium for Stirling at Los Alamos, ``I found Stirling to be insightful, creative, and really wonderfully unpretentious and easy-going in comparison to many individuals of comparable intellectual ability. I always treasured all those three qualities of him. I always felt I was learning from someone who confronted reality rather than pushed the equations.'' 
   
 \begin{figure}
  \begin{minipage}[t]{0.23\textwidth}
    \includegraphics[width=\textwidth]{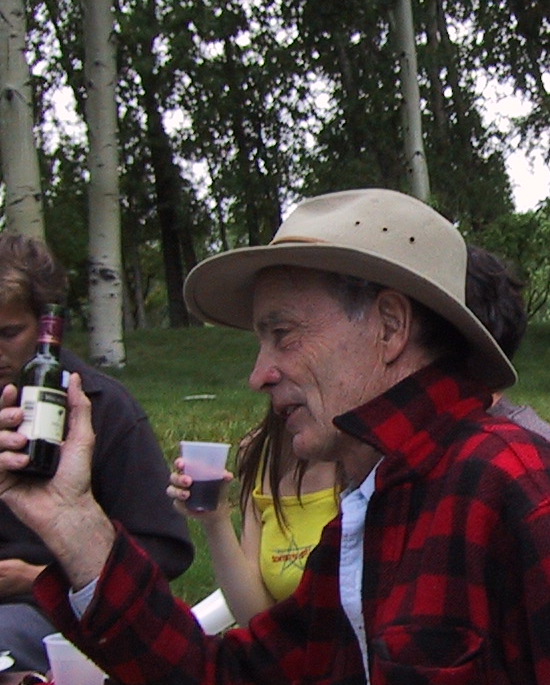}
    \caption{\footnotesize{Picnicking at the Aspen Center for Physics, 2001}}
    \label{fig:1}
  \end{minipage}
  \hspace{72pt}
  \begin{minipage}[t]{0.3\textwidth}
    \includegraphics[width=\textwidth]{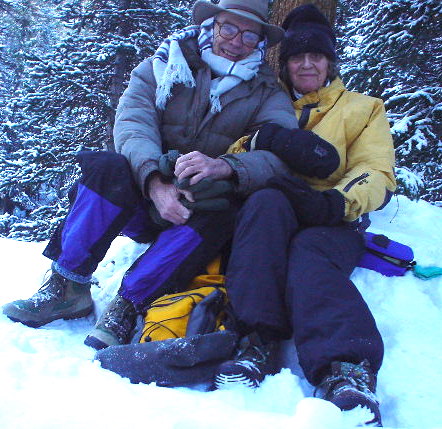}
    \caption{\footnotesize{Stirling and Rosie on winter hike to Cathedral Lake near Aspen, 2003.}}
    \label{fig:2}
  \end{minipage}
\vspace{-12pt}
\end{figure}

    Stirling was elected to the National Academy of Sciences in 1984.  The High Energy Astrophysics Division of the American Astronomical Society awarded him the Bruno Rossi Prize in 1990 ``in recognition of his seminal role in predicting the generation of  neutrinos in core collapse and elucidating the importance of the neutrinos for the dynamics 
and diagnostics of supernova explosions.''  Then the Franklin Institute ``selected as the recipient of the award of the John Price Wetherill Medal for 1994 Stirling A. Colgate for his fundamental contribution the the understanding of stellar collapse and supernova explosions.''   He received the 2006 Los Alamos medal from the Los Alamos National Laboratory in a citation beautifully 
summarizing his career:
 ``For leading the nuclear diagnostics of the nation's largest weapons test, conducted by Los Alamos; for negotiating the cessation of high-altitude and outer space nuclear tests, inspiring the inertial fusion and astrophysics programs at Livermore and Los Alamos, and contributing basic science to fusion ignition and burn, plasma confinement, and shock wave physics; for seminal work in supernovae and gamma ray bursts, recruiting leading weapons physicists through joint appointments in weapons design and astrophysics, and demonstrating by example that basic and applied science must be partners.''

  \begin{wrapfigure}[13]{R}{6cm}
\begin{center}
\vspace{-12pt}
\includegraphics[width=6cm]{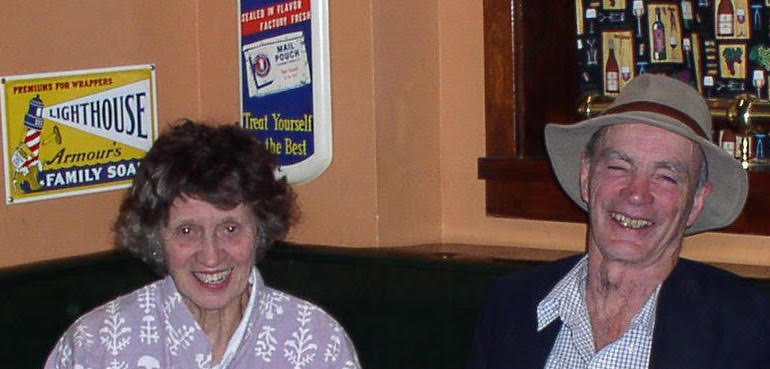}
\end{center}
\caption{\footnotesize{Stirling and Rosie, ca. 2005.}}
\label{2005}
\end{wrapfigure}

    Stirling was a regular participant over the years in the Aspen Center for Physics, enjoying the science, the people, and the mountains.  He was among a group of senior scientists at Los Alamos National Laboratory in the early 1980s under George Cowan, including Nicholas Metropolis, Herbert Anderson, Darragh Nagle, Peter Carruthers, and Richard Slansky, whose vision for an independent, trans-disciplinary scientific center would grow into the Santa Fe Institute in 1984.

   Scientifically active to the end, Stirling died on December 1, 2013 at his home in White Rock,  New Mexico. His wife, Rosemary Williamson -- always Rosie -- whom he married in 1947, passed away  April 19, 2018.  They shared a love of the outdoors, and will be remembered for their immense generosity.   They are survived by their son Arthur, daughter Sarah Chase, five grandchildren and seven great-grandchildren. Their son Hank Colgate died early, as did one grandchild.

\end{document}